\documentclass[prl,aps,twocolumn,showpacs,amsmath,amssymb]{revtex4}
\usepackage{graphicx}
\usepackage{dcolumn}
\usepackage{bm}
\usepackage{subfigure}
\usepackage{setspace}
\begin{document}
\title{Fermionization and bosonization of expanding 1D anyonic fluids}

\author{A. del Campo}
\email[Email address: ]{adolfo.delcampo@ehu.es}
% \affiliation{Institute for Mathematical Sciences, Imperial College London, SW7 2PE, UK}
% \affiliation{QOLS, The Blackett Laboratory, Imperial College London, Prince Consort Rd., SW7 2BW,UK}
\affiliation{Departamento de Qu\'\i mica-F\'\i sica, Universidad del
Pa\'\i s Vasco, Apartado 644, 48080 Bilbao, Spain}

\def\d{{\rm d}}
\def\la{\langle}
\def\ra{\rangle}
\def\om{\omega}
\def\Om{\Omega}
\def\vep{\varepsilon}
\def\wh{\widehat}
\def\tr{\rm{Tr}}
\def\da{\dagger}
\newcommand{\beq}{\begin{equation}}
\newcommand{\eeq}{\end{equation}}
\newcommand{\beqa}{\begin{eqnarray}}
\newcommand{\eeqa}{\end{eqnarray}}
\newcommand{\intf}{\int_{-\infty}^\infty}
\newcommand{\into}{\int_0^\infty}
\date{\today}
\begin{abstract}
The momentum distribution of an expanding cloud of one-dimensional hard-core anyons 
is studied by an exact numerical approach, and shown to become indistinguishable from that 
of a non-interacting spin-polarized Fermi gas for large enough times (dynamical fermionization). 
We also consider the expansion of one-dimensional anyons with strongly attractive short-range interactions suddenly released from a parabolic external potential, and find that momentum distribution approaches that of its dual system, the ideal Bose gas (dynamical bosonization). 
For both processes the characteristic time scales are identified, and the effect of the initial confinement is analyzed 
comparing the dynamics associated with both harmonic and hard-wall traps.
\end{abstract}
\maketitle
%
%

%{\it Introduction.}

The recent realization of ultracold atoms in tight-waveguides 
has spurred extensive studies of one-dimensional (1D) systems exhibiting Bose-Fermi duality \cite{CS99}.
The inextricable link between interactions and exchange statistics in 1D allows to 
relate the Bose gas with $\delta$-interactions with spin-aligned attractive Fermi gas with inverse coupling constant.
The upshot is that, in spite of the different symmetry of the many-body wavefunction, 
dual systems share many of its properties, including any local correlation function.

Paradigmatic examples are the 1D gas of spin polarized-fermions and impenetrable bosons 
(the so-called Tonks-Girardeau (TG) gas \cite{Girardeau60}) which have been prepared in the laboratory \cite{exp}. 
Similarly, the duality between non-interacting bosons and spin-polarized fermions 
with strongly attractive odd-wave interactions (fermionic Tonks-Girardeau (FTG) gas \cite{FTG}) 
has recently been explored.
A staggering manifestation of the BF duality is the dynamical fermionization 
in which the momentum distribution of the TG gas approaches during 
a 1D free expansion that of non-interacting fermions \cite{RM05,MinGan05}.
The reverse phenomenon has been predicted for FTG under the same dynamics, 
in which asymptotically the momentum distribution of the ideal Bose gas is exhibited \cite{GMin06}. 
Far from being limited to the Bose-Fermi mappings, 
the duality in lower dimension extends to particles with fractional statistics, 
whose realization with ultracold atoms has recently been proposed \cite{PFCZ01}.
Motivated by its application to transport in quantum Hall fluids, 
and the possibility of performing universal quantum computation \cite{DFNSS07}, 
the properties of 1D anyons are under current scrutiny. 
Particularly, several studies have been devoted to 1D anyons with short-range delta interactions (anyonic Lieb-Liniger model) 
\cite{Kundu99,BGO06,CM07,PKA07,Hao08} 
and the limiting cases of impenetrable anyons \cite{Girardeau06,SSC07,PKA08,SC08}, though much less is known about its dynamics.

In this letter, we shall consider the anyonic generalization of TG and FTG gases, 
and show that they do undergo dynamical ``dualization'' during their free time evolution.

\begin{figure}
\includegraphics[width=6cm,angle=0]{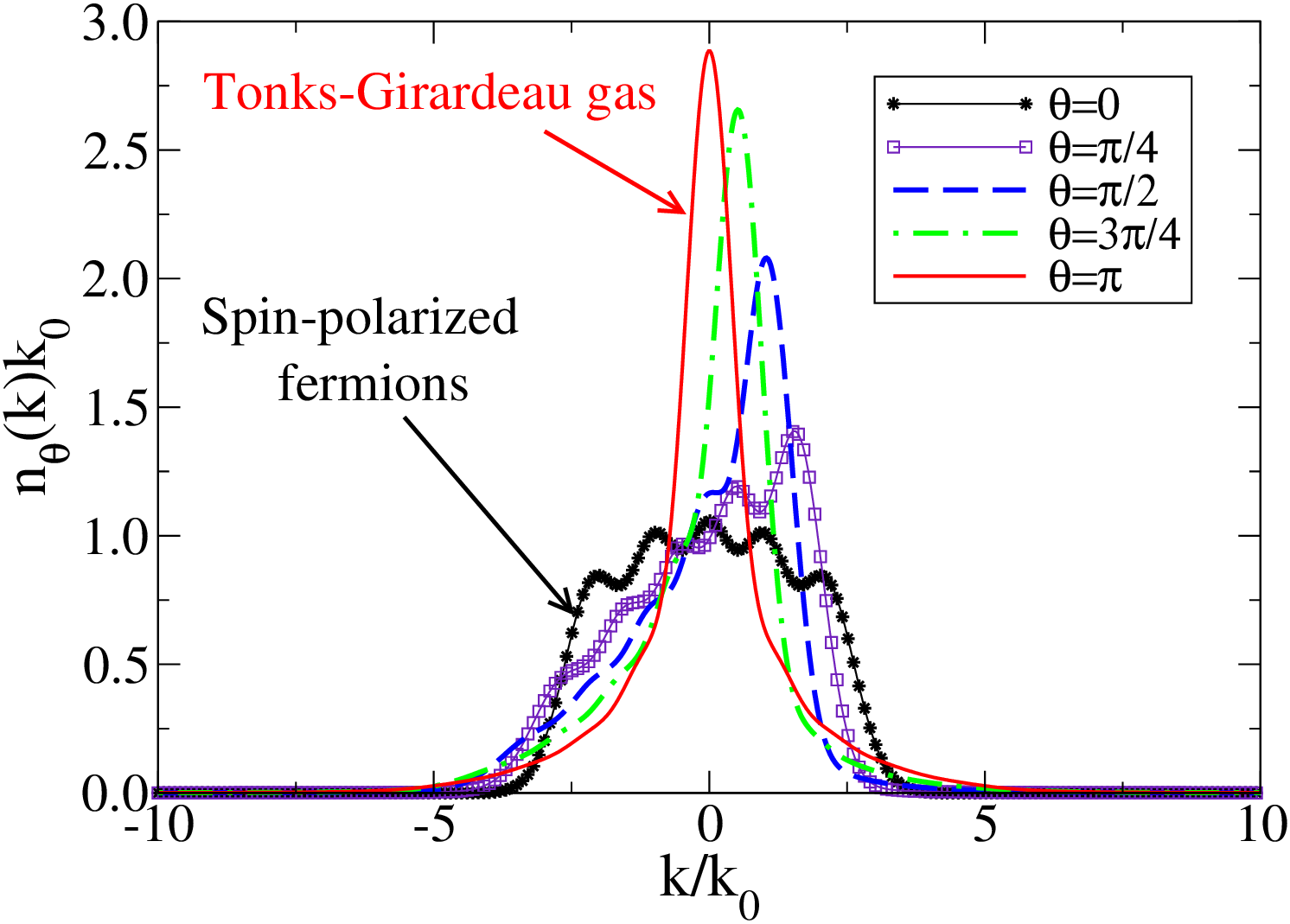}
\hspace{-.4cm}

\caption{\label{nkatgtrap} Momentum distribution of harmonically trapped  hard-core anyons, 
smoothly extrapolating between the bosonic Tonks-Girardeau gas ($\theta=0$) and spin-polarized non-interacting fermions ($\theta=\pi$), 
for $N=5$ particles ($k_0=\sqrt{m\om_0/\hbar}$, in all figures).} 
\end{figure}

{\it Hard-core anyonic gas.}
Strongly $\delta$-interacting anyons, at low enough densities may reach the so-called Tonks-Girardeau 
regime in which the particles become impenetrable. We assume that the gas of hard-core anyons (HCA) is initially confined in a harmonic trap of frequency $\om_0$, whose eigenstates are denoted by $\phi_{n}(x)$.
The description in such regime is simplified introducing a dual system, a spin polarized Fermi gas, whose ground-state wavefunction is a normalized Slater determinant \cite{Girardeau60},
$\Psi_{F}(x_{1},\dots,x_{N}) =\frac{1}{\sqrt{N!}}{\rm det}_{n,l=(0,1)}^{(N-1,N)}\phi_{n}(x_{l})$.
The wavefunction of the ATG gas is obtained from $\Psi_F$ by imposing the correct symmetry under permutation of particle
using the anyon-fermion (AF) mapping \cite{Girardeau06} 
$
\Psi_{HCA}(x_{1},\dots,x_{N})= \mathcal{A}_{\theta}^{\dagger}(\hat{x}_{1},\dots,\hat{x}_{N})\Psi_{F}(x_{1},\dots,x_{N}),\nonumber
$ 
where  
%
%\beqa
$\mathcal{A}_{\theta}=\prod_{1\leq j<k\leq N}e^{i\frac{\theta}{2}\epsilon(\hat{x}_{k}-\hat{x}_{j})}$
%\nonumber\eeqa
%
 is the one-parameter family of unitary operators which generalizes the well-known Bose-Fermi map \cite{Girardeau60} and $\epsilon(x)=1$ $(-1)$ if $x>0$ $(<0)$ and $\epsilon(0)=0$.
Varying the statistical parameter $\theta$, the map smoothly extrapolates between spin-polarized fermions ($\theta=0$), and the well-known 
bosonic Tonks-Girardeau gas ($\theta=\pi$) \cite{note1}. Note that  
$\mathcal{A}_{-\theta}=\mathcal{A}_{\theta}^{\dagger}=\mathcal{A}_{\theta}^{-1}$ and the expectation values coincide  $\la\Psi_{HCA},\Psi_{HCA}\ra=\la\Psi_{F},\Psi_{F}\ra$,
which imply that both dual systems share the same density profile \cite{Girardeau06} $\rho_{\theta}(x,t)= N\int\vert\Psi(x,x_{2},\dots,x_{N};t)\vert^{2} \d x_{2} \cdots\d x_{N}
=\sum_{n=0}^{N-1}\vert\phi_{n}(x,t)\vert^{2}$. However, its momentum distribution  
%
% \beqa 
% \label{nk}
$n(k)=(2\pi)^{-1}\int\!\!\int\d x\d ye^{ik(x-y)}\rho(x,y)$ 
%\eeqa 
%
is drastically different. For the spin polarized fermions, described by an Slater determinant wavefunction, the reduced single-particle density matrix (RSPDM) simply reads $\rho_{F}(x,y)=N\int\Psi_F(x,x_{2},\dots,x_{N})^{*}\Psi_F(y,x_{2},\dots,x_{N}) \d x_{2} \cdots\d x_{N}=\sum_{n=0}^{N-1}\phi_{n}^*(x)\phi_n(y)$ (normalized to $N$), whence it follows that $n_F(k)=\sum_{n=0}^{N-1}|\tilde{\phi}_n(k)|^2$ (the tilde denoting Fourier transform).
For the HCA gas, making use explicitly of the anyon-Fermi map, the orthonormality of the single-particle eigenstates, 
and the Laplace expansion of the determinant in the dual wavefunction, it is found that the RSPDM can be efficiently computed as
\beqa
\rho_{HCA}(x,y)=\sum_{l,n=0}^{N-1}\phi_l^*(x){\rm A}_{ln}(x,y)\phi_n(y),
\eeqa
 where ${\rm \bf A}(x,y)=({\bf P}^{-1})^T {\rm det}{\bf P}$
 and the elements of the matrix ${\bf P}(x,y)$ are 
\beqa 
P_{ln}=\delta_{ln}-(1-e^{-i\theta\epsilon(y-x)})
\epsilon(y-x)\int_{x}^{y}\d z \phi_{l}^*(z)\phi_n(z)\nonumber
\eeqa
a result which holds under time-evolution. This generalizes for any anyonic parameter, the result recently obtained 
by Pezer and Buljan \cite{PB07} for the Tonks-Girardeau gas ($\theta=\pi$). As shown in Fig. \ref{nkatgtrap}, an 
asymmetric momentum distribution results for any statistical parameter other than $\theta=0,\pi$ \cite{Hao08,SC08}. 
Moreover, for $\theta=0$, $P_{ln}=\delta_{ln}$ 
so the RSPDM becomes diagonal, and the familiar result for spin-polarized fermions is recovered.
In what follows we shall be interested in the expansion dynamics after suddenly switching off the confining potential at $t=0$.
We find that, as the time of evolution goes by, the momentum distribution of the HCA gas approaches that of its dual system, the spin-polarized ideal Fermi gas. 
%, which is suddenly turned off at $t=0$.
%
%%%%%%%%%%%%%%%%%%%%%%%%%%%%%%%%%%%%%%%%%%%%%%%%%%%%%%%%%%%%%%%%%%%%%%%%%%%%%
%%%%%%%%%%%%%%%%%%%%%%%%%%%%%%%%%%%%%%%%%%%%%%%%%%%%%%%%%%%%%%%%%%%%%%%%%%%%%%
%

%
%%%%%%%%%%%%%%%%%%%%%%%%%%%%%%%%%%%%%%%%%%%%%%%%%%%%%%%%%%%%%%%%%%%%%%%%%%%%%
\begin{figure}
\includegraphics[width=8.cm,angle=0]{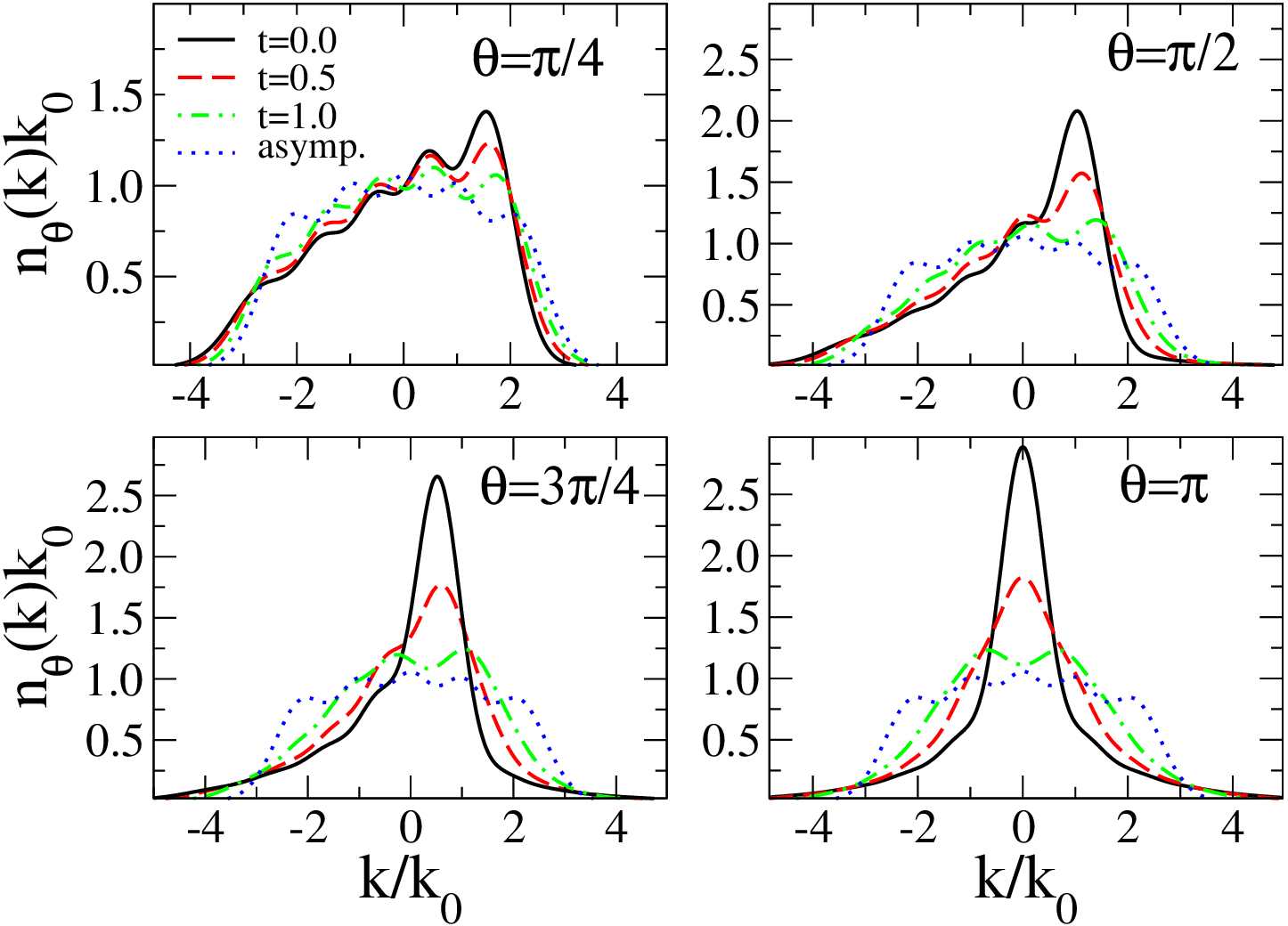}
\hspace{-.4cm}

\caption{\label{fermionization} Dynamical fermionization on the momentum distribution of hard-core anyons 
during a 1D free-expansion for different statistical parameters ($N=5$). Asymptotically, the momentum distribution of non-interacting spin-polarized fermions is obtained. The time of evolution is measured in units $\om_0^{-1}$ where $\om_0$ is the frequency of the initial harmonic confinement. 
} 
\end{figure}
%%%%%%%%%%%%%%%%%%%%%%%%%%%%%%%%%%%%%%%%%%%%%%%%%%%%%%%%%%%%%%%%%%%%%%%%%%%%%%
%
Provided that $\mathcal{A}_{\theta}$ is time-independent (the quantum statistics is preserved under time-evolution), 
the expansion dynamics of the manybody wavefunction is found by using in the anyon-fermion mapping.
%, the freely time-evolved single-particle states $\phi_n(x,t)$.
The self-similar evolution of the single-particle states according to the time-dependent Schr{\"o}dinger equation,
%\beqa
$i\hbar\frac{\partial\phi_n(x,t)}{\partial t}=\big[-\frac{\hbar^2}{2m}\frac{\partial^2}{\partial x^2}+\frac{m\om_0^2x^2\Theta(-t)}{2}\big]\phi_n(x,t)$,
%\eeqa
with $\Theta(t)$ being the Heaviside step function,
can be found exploiting the scaling law, $\phi_n(x,t)=\phi_n(x/b(t),0)e^{imx^2\dot{b}/2b\hbar-iE_n\tau(t)/\hbar}/\sqrt{b(t)}$ where the scaling factor $b(t)$ is the solution of the differential equation $\ddot{b}+\om^2(t)b=\om_{0}^2/b^3$ satisfying $b(0)=1$ and $\dot{b}(0)=0$, $E_n=\hbar\om_0(n+1/2)$, and $\tau(t)=\int_{0}^tdt'/b^2(t')$ \cite{PZ98}. For $\om(t)=\om_0\Theta(-t)$ it follows that 
$b(t)=\sqrt{1+\om^2_0t^2}$, and therefore, for all $t\gg\om_0^{-1}$ the expansion becomes ballistic \cite{OS02,MinGan05}. 
During the expansion a dynamical fermionization occurs, see Fig. \ref{fermionization}, by means of which the momentum distribution tends to that of the non-interacting spin-polarized Fermi gas, $n_F(k)=\sum_{n=0}^{N-1}|\tilde{\phi}_n(k)|^2$ where $\tilde{\phi}_n(k)=(-1)^n(2x_0^2/\pi)^{1/4}
e^{-k^2x_0^2}H_n(\sqrt{2}kx_0)$, $H_n$ are the Hermite polynomials and $x_0=\sqrt{\hbar/m\om_0}$. 
Indeed, the stationary phase method (SPM) allows to find the asymptotic form of the momentum distribution for $t\rightarrow\infty$ \cite{MinGan05}, $n_{AHA}(k)\sim|\om_0/\dot{b}|n_F(\om_0 k/\dot{b})\sim n_F(k)$. Intuitively, as the gas expands, the particles stop to interact, and asymptotically one find the distribution of quasi-momenta, which are integrals of motion.
The fermionization time scale can be approximated as $t_{F}(\theta)\approx \sin(\theta/2)/N\om_0$, dependent on the statistical parameter $\theta$ and  decreasing with the number of particles $N$.

{\it Expansion from a box.}
The scaling law is specific of the harmonic confinement 
and we may rightly wonder how the subsequent time-evolution differs from the expansion from other type of traps.
In what follows we shall compare the previous results with the expansion from a box-like trap \cite{DM05}. The main difference is that at variance with the parabolic potential the expansion from a hard-wall confinement is not self-similar,
as the space-time dynamics already exhibits a rich transient structure at the single-particle level. 
For a hard-wall confinement of width $L$, the well-known orthonormal eigenstates read $\varphi_j(x)=(2/L)^{1/2}\sin(k_j x)\chi_{[0,L]}(x)$ ($k_j=j\pi/L$, $j\in\mathbb{N}$, $\chi_{[0,L]}(x)$ is the characteristic function in the interval $[0,L]$, and the corresponding energy eigenvalues are $E_j=\hbar^2 k_j^2/2m$).
The exact time-evolution of the single-particle eigenstates of a hard-wall trap in a 1D free expansion $\varphi_{j}(x,t)$
 can be found using the superposition principle $\varphi_j(x,t)=\int_{-\infty}^{\infty}dx'K_0(x,t,x',t=0)\varphi_j(x',0)$ where the free propagator 
$K_0(x,t,x',0)=(m/2\pi i\hbar t)^{1/2}{\rm exp}(im(x-x')^2/2\hbar t)$. After performing the integral explicitly, 
$\varphi_{j}(x,t) =
\frac{1}{i\sqrt{2L}}\sum_{\alpha=\pm1}\alpha
\big[e^{i\alpha k_{j}L}M(x-L,\alpha k_{j},\tau)
-M(x,\alpha k_{j},\tau)\big]$ 
where $\tau=\hbar t/m$, the Moshinsky function is defined as $M(x,k,t)=e^{i\frac{x^{2}}{2t}}w\big[-\frac{1+i}{2}\sqrt{t}
\left(k-\frac{x}{t}\right)\big]/2$ \cite{Moshinsky52}, and the Faddeyeva function is related to the 
complementary error function $w(z)=e^{-z^{2}}{\rm Erfc}(-iz)$ \cite{AS65}. For all $j>1$, the density profile $|\varphi_{j}(x,t)|^2$ bifurcates in two main branches moving with $\pm k_j$ after a critical time $t_j=mL^{2}/2j\pi\hbar$ reflecting the underlying bimodal momentum distribution \cite{spnk}. For the free expansion of a TG gas the transition to the ballistic regime  sharply takes place at $t_N$ \cite{DM05}. 
%
%%%%%%%%%%%%%%%%%%%%%%%%%%%%%%%%%%%%%%%%%%%%%%%%%%%%%%%%%%%%%%%%%%%%%%%%%%%%%
\begin{figure}
\includegraphics[width=8.cm,angle=0]{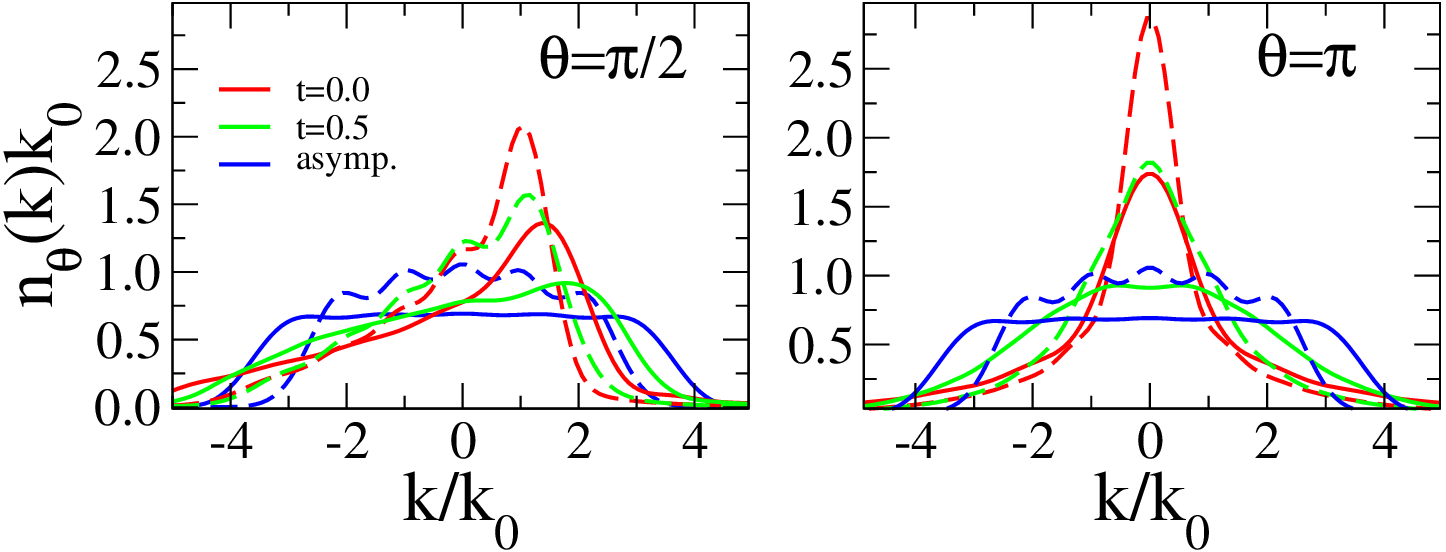}
\hspace{-.4cm}
\caption{\label{box} Effect of the confining potential. The momentum distribution of the 
anyonic Tonks-Girardeau gas released from the box (solid lines) broadens with respect to the harmonic case (dashed lines), 
and undergoes fermionization in a shorter time scale ($N=5$). Note that 
$\om_0$ and $L$ are chosen for the total energy of the gas to be the same in both traps (see text).} 
\end{figure}
%%%%%%%%%%%%%%%%%%%%%%%%%%%%%%%%%%%%%%%%%%%%%%%%%%%%%%%%%%%%%%%%%%%%%%%%%%%%%%
%
To compare the expansion from both harmonic and hard-wall traps we impose the particle number $N$ and total energy of the AHC gas to be the same. 
For the many-body ground-state, the following relation must hold between the frequency $\om_0$ and the length of the box, $L$: 
$\om_0=\hbar \pi^{2}(N+1)(2N+1)/6mL^{2}N$. Figure \ref{box} shows  how the momentum distribution is broadened with the initial hard-wall confinement with respect to the harmonic case, due to the dispersion relation $E_n\propto n^2$ and the presence of higher momentum components. 
The reference case for $\theta=\pi$, the bosonic TG gas, presents a sharper peak at $k=0$ when prepared in the later trap. 
Note also how asymptotically the quasi-momentum distribution is already flat (for $N=5$) for the box confinement, while in the harmonic trap case keeps an oscillating profile, both of which are eventually mapped into the coordinate density profile in a $t_F$ time scale.

{\it Anyonic attractive Tonks-Girardeau gas.}
The anyonic attractive Tonks-Girardeau (AATG) gas has recently been introduced in analogy with the Fermionic Tonks-Girardeau gas (FTG) \cite{Girardeau06}. 
The FTG results when a spin-polarized Fermi gas is driven through a p-wave Feschbach resonance inducing strongly attractive short range odd-wave interactions \cite{BEG05}. Indeed, in a given sector $x_1<x_2$, the resulting pseudopotential consists of a hard-core wall at $x_{12}=x_1-x_2=0$ and a square well of depth $V$ and width $a$ in the limit $a\rightarrow0$, $V\rightarrow\infty$, keeping $V=(\hbar\pi/a)^2/4m$.  The dual system of the FTG is simply the ideal Bose gas, whose ground state is simply described by a Hartree product  $\Psi_{B}(x_{1},\dots,x_{N})=\prod_{n=1}^{N}\phi_{0}(x_n)$, $\phi_{0}$ being the single particle ground state of the external potential. This motivates, the study of the 1D anyons with such infinitely attractive pseudopotential -the AATG gas-, whose ground state wavefunction is given by the anyon-boson mapping  
%
%\beqa
$\Psi_{AATG}(x_{1},\dots,x_{N})=\prod_{1\leq j<k\leq N}\epsilon(\hat{x}_{k}-\hat{x}_{j})e^{-i\frac{\theta}{2}\epsilon(\hat{x}_{k}-\hat{x}_{j})}\Psi_{B}(x_{1},\dots,x_{N})$,
%\eeqa 
where the mapping holds under any unitary time evolution \cite{Girardeau06}. 
%
%%%%%%%%%%%%%%%%%%%%%%%%%%%%%%%%%%%%%%%%%%%%%%%%%%%%%%%%%%%%%%%%%%%%%%%%%%%%%
\begin{figure}
\includegraphics[width=6cm,angle=0]{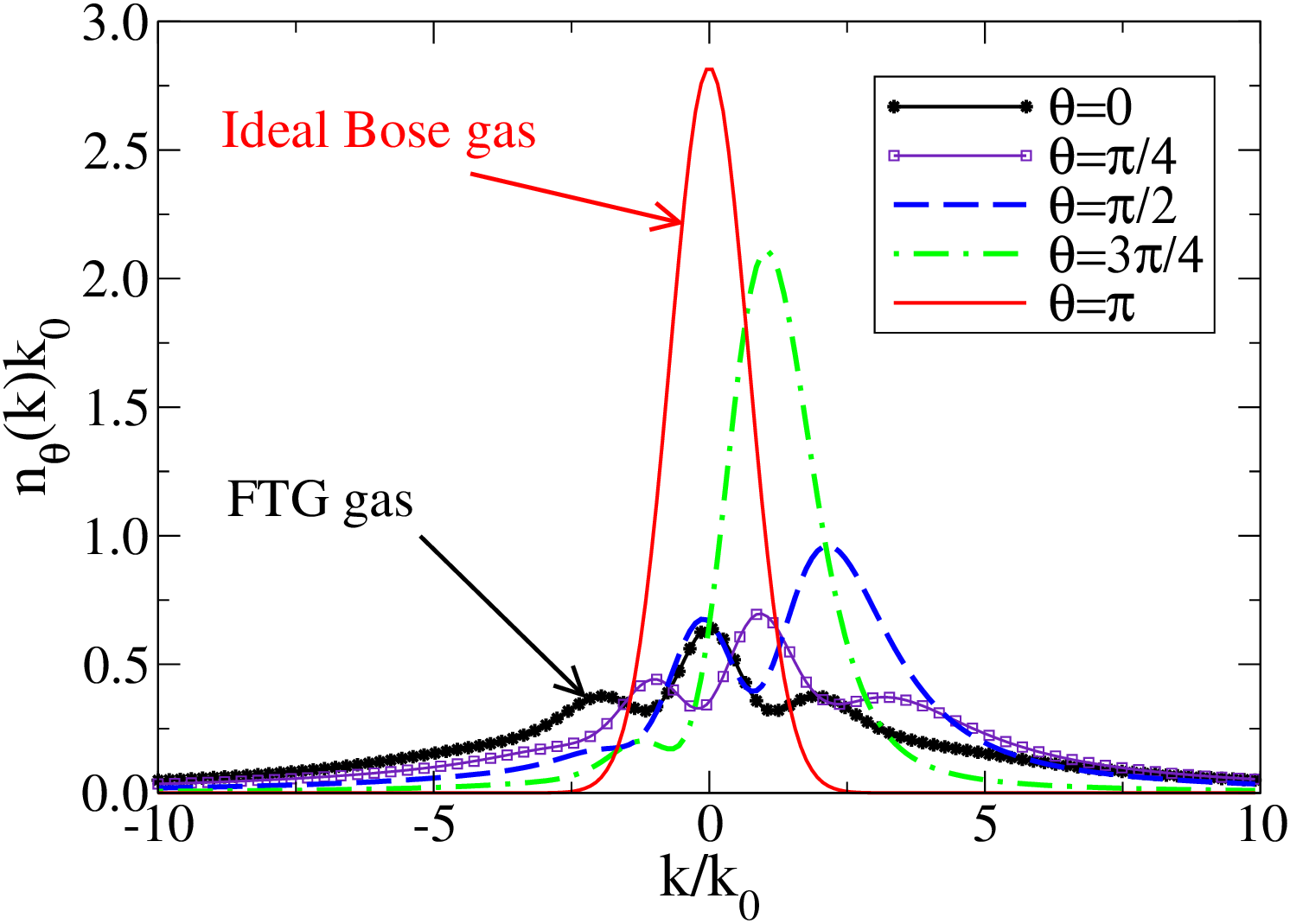}
\hspace{-.4cm}
\caption{\label{nkaatg} Momentum distribution of attractive hard-core anyons, 
smoothly extrapolating between the spin-polarized fermionic Tonks-Girardeau gas ($\theta=0$) and the non-interacting Bose gas ($\theta=\pi$) 
for $N=5$ particles in a parabolic external potential.} 
\end{figure}
%%%%%%%%%%%%%%%%%%%%%%%%%%%%%%%%%%%%%%%%%%%%%%%%%%%%%%%%%%%%%%%%%%%%%%%%%%%%%%
%
%
A one-parameter family of RSPDM for the AATG results from direct computation in the line of \cite{BEG05}, 
\beqa
\label{AATG}
\rho_{AATG}(x,y)
%&=&
% N\phi_0(x)\phi_{0}^{*}(y)\int\big[\prod_{n=2}^{N}\epsilon(x-z_{n})\epsilon(y-z_{n})
% e^{-\theta[\epsilon(y-z_{n})-\epsilon(x-z_{n})]}\big]\nonumber\\
% &\times& \prod_{1<j<k}^{N}|\epsilon(z_j-z_k)\epsilon(y-z_{n})|^2\phi_0(z_k)\phi_{0}^{*}(z_)\nonumber\\
&=&N\phi_0(x)\phi_{0}^{*}(y)[P_{0}^{\theta}(x,y)]^{N-1},
\eeqa
where 
\beqa P_{0}^{\theta}(x,y)\!\!&=&\!\!\int\d z\epsilon(y-z)\epsilon(x-z)e^{-i\frac{\theta}{2}[\epsilon(y-z)-\epsilon(x-z)]}|\phi_0(z)|^2\nonumber\\
&=&1-[1+e^{-i\theta\epsilon(y-x)}]\bigg|\int_{x}^y\d z|\phi_0(z)|^2\bigg|.
\eeqa
%
%assuming $x<z$ without loss of generality. 
In particular, for the harmonic confinement, 
%
%\beqa
%\label{P0}
$P_{0}^{\theta}(x,y)=1-\frac{1+e^{-i\theta\epsilon(y-x)}}{2}|{\rm Erf}(y/x_0\sqrt{b})
-{\rm Erf}(x/x_0\sqrt{b})|$.
%\eeqa
%
The first term in Eq. (\ref{AATG}), $N\phi_0^{*}(x)\phi_{0}(y)$, corresponds to the RSPDM of the dual non-interacting Bose system. 
Moreover, the momentum distribution $n_{\theta}(k)$ can be obtained by the usual double Fourier transform. 
Figure \ref{nkaatg} shows $n_{\theta}(k)$ for a gas of AATG in a harmonic trap as a function of the statistical parameter $\theta$.
The ideal gas distribution $N|\tilde{\phi}_0(k)|^2$ is recovered from the general expression for $\theta=\pi$, whereas the FTG gas is obtained at $\theta=0$. However, for any other value of the statistical parameter $\theta$ the momentum distribution of the trapped system is asymmetric as a result of the anyonic permutation symmetry.
%
%%%%%%%%%%%%%%%%%%%%%%%%%%%%%%%%%%%%%%%%%%%%%%%%%%%%%%%%%%%%%%%%%%%%%%%%%%%%%
\begin{figure}[t]
\includegraphics[width=8.cm,angle=0]{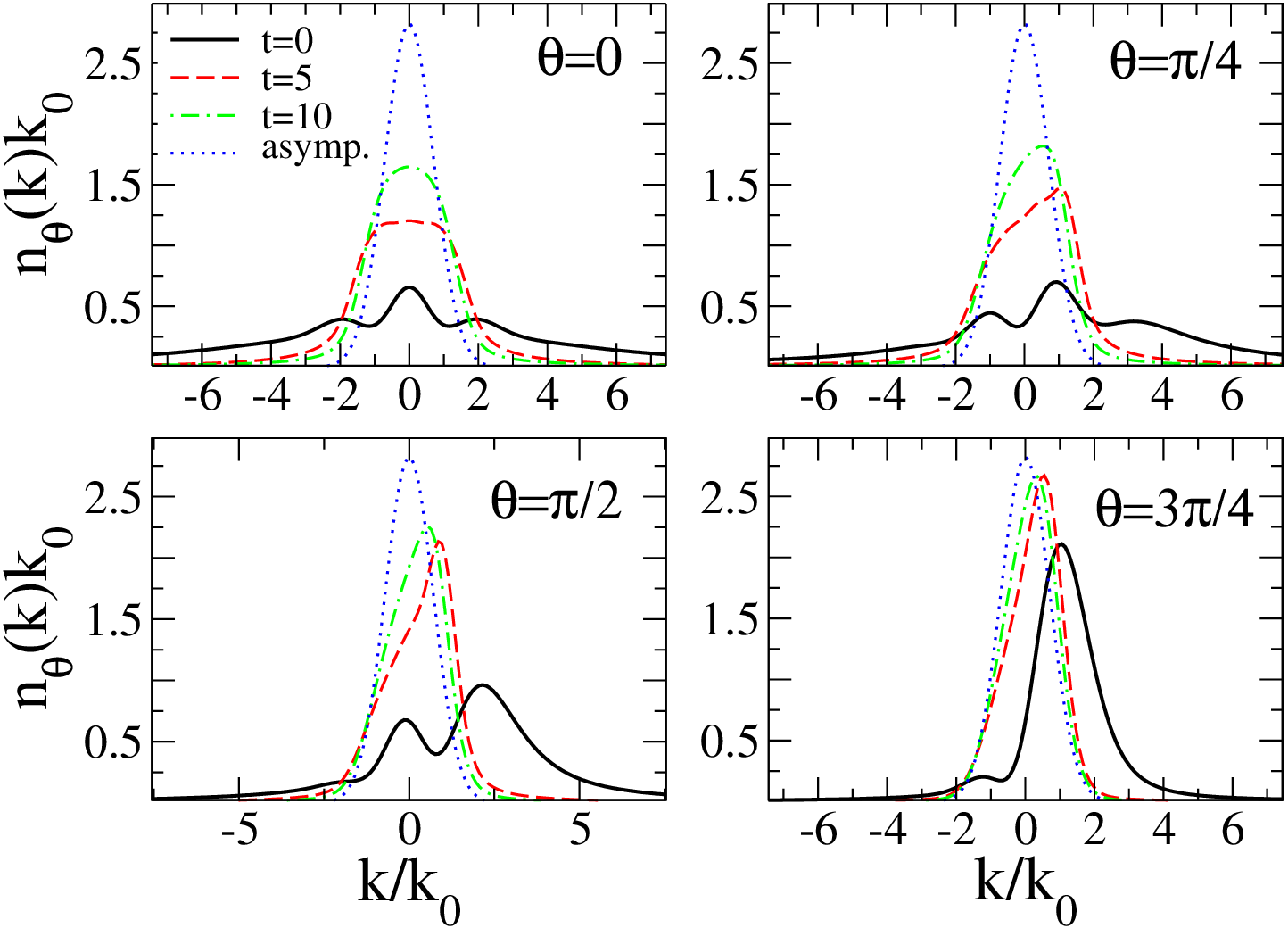}
%\hspace{-.4cm}

\caption{\label{bosonization} Dynamical bosonization on the momentum distribution of attractive hard-core 
anyons during a 1D free-expansion from a harmonic trap for different statistical parameters 
($N=5$). Asymptotically the momentum distribution of an ideal Bose gas is recovered.
} 
\end{figure}
%%%%%%%%%%%%%%%%%%%%%%%%%%%%%%%%%%%%%%%%%%%%%%%%%%%%%%%%%%%%%%%%%%%%%%%%%%%%%%
%
We shall next describe the one dimensional expansion dynamics of a given AATG, following a sudden switch-off  of the hard-wall trap.
Figure \ref{bosonization} shows how as the system expands in free space, the AATG  gas undergoes bosonization reaching asymptotically the (symmetric) momentum distribution of the ideal Bose gas, $n_{B}(k)= N|\tilde{\phi}_0(k)|^2$. 
For $t\rightarrow\infty$ the argument of each of the complementary error functions in $P_{0}^{\theta}(x,y)$ tends to vanish, 
and hence, ${\rm Erf}(0)=0$, and $P_0^{\theta}(x,y)=1$. It follows that $\rho_{AATG}(x,y)\sim N\phi_0^{*}(x)\phi_{0}(y)$, 
meaning that any one-particle observable, both local and non-local, becomes identical for both the AATG and ideal Bose gases.
The SPM leads for the the asymptotic momentum distribution $n_{AATG}(k)\sim|\om_0/\dot{b}|n_B(\om_0 k/\dot{b})\sim n_B(k)$.
For $N,\om_0 t\gg 1$, the off-diagonal correlations vary exponentially as 
$\rho_{AATG}(x,y,t)\sim\rho_B(x,y,t)\exp[\exp(-i\theta\epsilon(y-x)/2)N\cos(\theta/2)|y-x|/\om_0 t\sqrt{\pi}x_0)]$.
Hence, the bosonization time scale becomes $t_{B}(\theta)\approx N\cos(\theta/2)\om_0^{-1}$, this is, $t_{B}(\theta)\approx N^2t_{F}(\pi-\theta)$. 
Should we have considered the expansion from a hard-wall trap, the characteristic time scale would be 
$t_{B}(\theta)\approx NmL^2\cos(\theta/2)/2\pi\hbar$.

In conclusion, we have shown that 1D hard-core anyons 
undergo dynamical fermionization during a free time-evolution, namely, 
the momentum distribution of the gas approaches that of spin-polarized non-interacting fermions.
We have further shown that the complementary dynamical process can take place:
for anyons with strongly attractive short range interactions (the anyonic attractive Tonks-Girardeau gas),  
  the freely time-evolving momentum distribution tends to that of an ideal Bose gas. In both cases 
the momentum distribution becomes symmetric, as the role of the anyonic permutation symmetry diminishes.  
One may expect that the recently developed methods for the dynamics of a bosonic Lieb-Liniger gas \cite{LLdyn}, 
could equally be generalized for fractional statistics allowing to study 
the expansion dynamics for finite interactions \cite{Sutherland98}.

The author acknowledges discussions with J. G. Muga, I. L. Egusquiza, and E. Ya. Sherman. 
This work has been supported 
by Ministerio de Educaci\'on y Ciencia (BFM2003-01003), 
and UPV-EHU (00039.310-15968/2004) and 
the Basque Government (BFI04.479).

%\section*{References}


\begin{thebibliography}{10}
\bibitem{note1} We follow the notation of \cite{Girardeau06} for the statistical parameter $\theta$, 
related to the $\kappa$-parameter in \cite{SC08} by $\theta=\pi(1-\kappa)$.
\bibitem{spnk} Note that $\vert\tilde{\phi}_n(k)\vert^{2}=
\frac{2\pi n^2}{L^3}[1-(-1)^n\cos(kL)]/(k^2-k_{n}^{2})^2$, and has 
two main peaks at $k=\pm k_n$ for all $n\in\mathbb{N}$ but the ground state, 
for which $\vert\tilde{\phi}_0(k)\vert^{2}$ is a bell-shaped distribution centered at $k=0$.
\end{thebibliography}
\end{document}